\documentclass[unnumsec,webpdf,contemporary,large]{oup-authoring-template}%

\graphicspath{{Fig/}}

\theoremstyle{thmstyleone}%
\theoremstyle{thmstyletwo}%
\theoremstyle{thmstylethree}%

\usepackage{soul}
\usepackage{graphicx}
\usepackage{color}
\definecolor{sara}{rgb}{0.133, 0.545, 0.133}

\definecolor{walter}{rgb}{0.133, 0.133, 0.545}

\begin{document}

\journaltitle{...}
\DOI{DOI HERE}
\copyrightyear{2023}
\pubyear{2023}
\access{Advance Access Publication Date: Day Month Year}
\appnotes{Paper}

\firstpage{1}


\title[locuaz]{Locuaz: an in-silico platform for antibody fragments optimization}

\author[1,2,$\ast$]{German P. Barletta\ORCID{0000-0002-0806-0812}}
\author[1]{Rika Tandiana\ORCID{0000-0000-0000-0000}}
\author[1,3]{Miguel Soler\ORCID{0000-0002-5780-9949}}
\author[1,4\#,$\ast$]{Sara Fortuna\ORCID{0000-0000-0000-0000}}
\author[1,\#,$\ast$]{Walter Rocchia\ORCID{0000-0000-0000-0000}}

\authormark{Barletta et al.}

\address[1]{\orgdiv{CONCEPT}, \orgname{Istituto Italiano di Teconologia}, \state{Genova}, \country{Italy}}
\address[2]{\orgname{The Abdus Salam International Centre for Theoretical Physics (ICTP)}, \state{Trieste}, \country{Italy}}
\address[3]{\orgdiv{Dipartimento di Scienze Matematiche, Informatiche e Fisiche (DMIF)}, \orgname{University of Udine}, \state{Udine}, \country{Italy}}
\address[4]{now at Cresset, New Cambridge House, Bassingbourn Road, Litlington, Cambridgeshire, SG8 0SS, UK}

\address[\#]{Contributed equally to this work}


\received{Date}{0}{Year}
\revised{Date}{0}{Year}
\accepted{Date}{0}{Year}



\abstract{
\textbf{Motivation:} Engineering high-affinity binders targeting specific antigenic determinants remains a challenging and often daunting task, requiring extensive experimental screening. Computational methods have the potential to accelerate this process, reducing costs and time, but only if they demonstrate broad applicability and efficiency in exploring mutations, evaluating affinity, and pruning unproductive mutation paths.
\newline
\textbf{Results:} In response to these challenges, we introduce a new computational platform for optimizing protein binders towards their targets. The platform is organized as a series of modules, performing mutation selection and application, molecular dynamics (MD) simulations to sample conformations around interaction poses, and mutation prioritization using suitable scoring functions. Notably, the platform supports parallel exploration of different mutation streams, enabling in silico high-throughput screening on HPC systems. Furthermore, the platform is highly customizable, allowing users to implement their own protocols.
\textbf{Availability and implementation:} the source code is available at \url{https://github.com/pgbarletta/locuaz} and documentation is at \url{https://locuaz.readthedocs.io/}
\newline
\textbf{Contact:} walter.rocchia@iit.it, sara.fortuna@cresset-group.com, pbarletta@gmail.com
\newline
{\textbf{Supplementary information:} available at \url{https://github.com/pgbarletta/suppl_info_locuaz}}
}


\maketitle
\section{Introduction}

The Antibody (Ab) discovery process, a critical aspect of biotherapeutics development, relies on the identification of one or more starting candidates, for instance via phage display, yeast display, and hybridoma technology. The affinity maturation follows through several steps of either site-directed mutagenesis, directed evolution, or deep mutational scanning which allow assessing the impact of specific mutations on Ab function and stability (\cite{kennedy2018monoclonal}). The goal is to obtain an Ab with optimal stability, high affinity and specificity toward the target. 
\textit{In-silico} approaches can be extremely useful in this context, especially when structural data of the target are available. They include knowledge-based methods, trained on sequence and structure databases, physics-based methods, and hybrid approaches, all aiming at mimicking the details of protein-protein interactions at a lower cost and a shorter time (\cite{sormanni2018third}).

Many popular empirical methods belong to the Rosetta suite(\cite{rosettantibody2009}). For instance, Rosetta RAbD, an iterative protocol generating redesigned mutants and samples conformations via a Monte Carlo (MC) scheme using a specific energy definition, led to the discovery of an Ab with a Kd \(\sim \)20nM (\cite{rosettantibody2018}). This result is comparable to those of experimental methods, which very rarely deliver Abs with pM binding affinity.

Physics-based approaches often rely on MD simulations for conformational sampling. Among them, a protocol based on MD coupled with a Metropolis Montecarlo scheme using energies calculated via MMPBSA to accept sampled conformations led to an Ab with Kd \(\sim \)0.5nM (\cite{buratto2022silico}). 
MD+FoldX is another method where the interaction energies are calculated exclusively with FoldX and it has been used to optimize antibodies against the SARS-Cov-2 receptor binding domain. (\cite{barnes2022})
In other approaches, the conformations generated via MD were selected with an acceptance criterion based on a consensus score among multiple scoring functions giving raise to an evolutive optimization protocol for nanobodies and peptides, reaching affinities comparable to those of experimental techniques (\cite{parcesoler2019,parce2021}). A later implementation exploiting replica-exchange MD allowed the \textit{de novo} optimisation of Ab fragments (\cite{soler2023replica})). 

%
%
Among the most advanced examples of knowledge-based methods, certainly those based on machine learning deserve to be mentioned, such as generative methods. In a recent effort, 1 million Abs against the HER 2 growth factor receptor have been generated in a zero-shot fashion, without any training sample binding HER 2 or one of his homologs. From these designs, 421 binders were experimentally validated and three displayed stronger binding than trastuzumab, the antibody binder licensed as a drug (\cite{absci2023}). 
Along the same lines, a deep learning based method trained on $10^4$ Ab variants of the antiHer2 Ab Trastuzumab, was used to pick among $10^8$ mutants achieving affinities of Kd 0.1-10nM, thus comparable to that of the original Ab they were derived from (\cite{mason2021optimization}), Language models too have been used to optimise Abs with Kd$<$1nM  (\cite{Hie2023}).

Overall, deep-learning based methods are very promising but need a very large training set of experimental data, which might not always be available.
These results highlight that \textit{in-silico} methods could be key to avoid massive experimental costs. Moreover, the variety of available protocols and of their intermediate steps allows to tailor each phase of the process, such as the mutation strategy or the affinity estimation criteria.

We propose here \emph{locuaz}: a flexible, python-based platform whose primary goal is to optimise the binding affinity of an Ab (fragment) towards its target. In an evolutionary framework, \emph{locuaz} mutates the candidate binder, samples conformations via MD and scores the affinity of the new complex. Each of these steps can be customized by the user. Importantly, the entire mutation pipeline allows the concurrent generation of different mutation lineages in parallel. The platform can be containerized and run through different job scheduling systems, such as SLURM or PBS. It was constructed using a modular approach both by selecting features from existing Ab evolutionary protocols and by developing new ones into a unique platform, in a way to facilitate updates, customizations and extensions.

\section{Methods}\label{metoditos}

\begin{figure*}[!ht]%
\centering
\includegraphics[width=1.\textwidth]{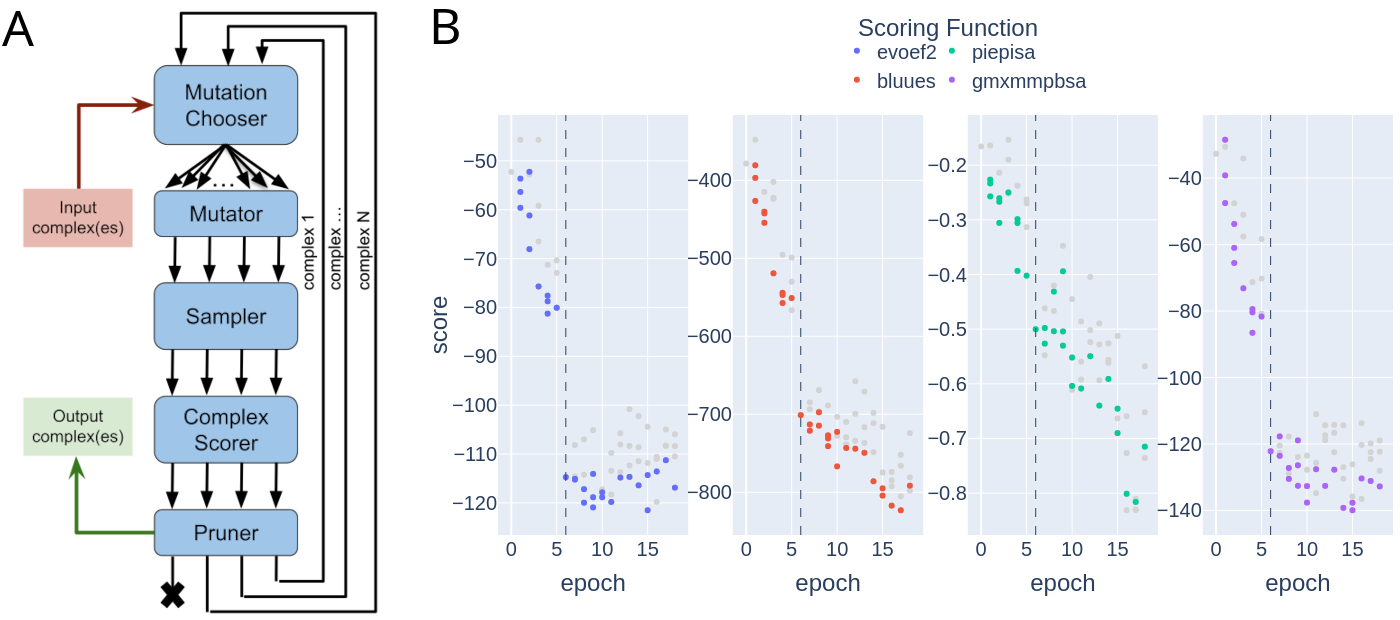}
\caption{\textbf{A} User configurable Blocks of the protocol.
\textbf{B} Protocol output along the optimisation of an anti-p53 antibody fragment: Average scores for all complexes generated along the protocol. Gray markers correspond to complexes that were not selected for the next epoch. Dotted line marks the first epoch of the unrestrained optimization, showing that the lifting of restrains allowed the interface to find a lower energy interaction.}\label{fig1}
\begin{picture}(0,0)
\end{picture}
\end{figure*}

As shown in Figure \ref{fig1}A, the workflow of \emph{Locuaz} is organized in functional units called Blocks, namely the \textbf{Mutation Chooser}, which determines which mutation has to be applied, the \textbf{Mutator}, which actually applies the mutation to the structure, the \textbf{Sampler}, which performs MD simulation to relax the system, the \textbf{Complex Scorer}, which evaluates the effect of the current mutation on the binding affinity, and finally the \textbf{Pruner}, which decides whether a lineage should be stopped because it is not promising or not. These units were envisioned to support the interchangeable use of built-in tools or of third-party external programs for each task. For example, \emph{Locuaz} currently supports 9 different \textbf{Scorers}, 
2 possibilities for the input topologies (Amber and GROMACS), GROMACS as MD engine used by the \textbf{Sampler}, 
3 different \textbf{Pruners} and a highly-configurable \textbf{Mutation Chooser}. Different flavors of each block can be selected and combined at will, and new ones can be added, leading to a remarkable protocol flexibility.

The optimization process starts from a putative binder/target complex and requires as input the selection of the portion of the binder to be optimised. The workflow then envisions the identification of multiple mutations by the \textbf{Mutation Chooser} block. The mutation sites can be one or many, can be either completely random, or guided by biological knowledge, or by physically-based methods, such as MMPB(GB)SA, which are used to recognize which residues are contributing less to the binding affinity (\cite{mmpbsa2021}).
The choice of the new amino acid is also configurable.
The user can choose from different amino acid probability schemes, or assign each amino acid a custom probability.
The \textbf{Mutation Chooser} also allows custom grouping of amino acids
in order to force the choice of the amino acid to be within a certain category of amino acid like "polar" or "aliphatic", etc.
More information is available on the documentation.

Mutations are performed by any of the currently available \textbf{Mutators} (\cite{tandiana2024, parcesoler2018a}). 

After a mutation is applied, the \textbf{Sampler} performs MD simulations using the GROMACS engine (\cite{gromacs2015}) by adopting either GROMACS topologies, or those from Amber's Tleap (\cite{Salomon2013}), allowing the user to include non-standard residues, organic small molecules and ions. To maximize the throughput, all available GPU and CPU resources are pooled and the  simulations are  distributed among parallel branches (as many as the user requested) and are queued up for concurrent execution following a Producer/Consumer strategy where, in case the number of requests cannot be fulfilled by the available resources, a "first come, first served" criterion is adopted.

Then, the sampled complex configurations are scored and their scores averaged. 
The score calculations are also dispatched to the available resources, as previously described, to optimize the overall efficiency.

Finally, the user-selected \textbf{Pruner} compares the scores of the mutated binders against the original ones and selects the subset of mutated binders that are deemed to improve affinity. If a single scoring function is used, one possibility is to adopt a Monte Carlo-based Pruner (\cite{Fortuna2017}), giving a chance also to mutations that are predicted to lead to slightly lower affinity. If, conversely, multiple scoring functions are employed, a consensus criterion \textbf{Pruner} can be used to decide which mutation \textbf{branch} is to be continued(\cite{parcesoler2019}).
In case no mutation is retained, the original binders are reused to generate a new set of mutants.

Each cycle of this protocol is called an \textbf{epoch}, while simultaneous mutations give rise to different \textbf{branches}. A typical optimization process involves many epochs, as depicted on Figure \textbf{1B}. At each epoch a user-defined number of branches are generated, effectively forming a Directed Acyclic Graph (DAG). The width of this DAG, i.e. the number of active branches, is managed according to the available computational resources. In Locuaz we included two methods: \textbf{variable-width DAG}, suitable when considerable computational resources are available, and \textbf{constant-width DAG}, in which the creation of new branches is limited by a previously selected constant width.

\section{Application to the TWIST1 system}\label{example}
TWIST1 is a transcription factor which promotes the MDM2-mediated degradation of p53, one of the main tumor suppressors, by interacting with its binding site on p53, known as "Twist-box". 
A llama nanobody (VHH) shown to interfere with TWIST1:p53 interaction by binding to p53 with moderate binding affinity was modelled by homology and then docked onto p53
(\cite{twist2022}). 

In order to improve the p53:VHH affinity, the complex was submitted to 5 epochs of optimization using positional restraints between the binder and the target. Then, 250ns of NPT were carried out to confirm the improved stability. The most representative structure was chosen after clustering, and used to start another 13 epochs of unrestrained optimization for a total of 18 epochs. The optimisation process was monitored by plotting the 4 selected scoring functions along the epochs (Figure \ref{fig1}.B). In this particular example the \textbf{Mutator Chooser} was \textit{SPM4i}, the \textbf{Sampler} used \textit{GROMACS} topologies, the \textbf{Scorers} were \textit{EvoEF2}, \textit{BLUUES}, \textit{PIEPISA} and \textit{gmxMMPBSA} and the \textit{Consensus Threshold} \textbf{Pruner} was used to select the best candidates.
This example, as well as further details on the specific features of each \textbf{Block}, can be found at \texttt{https://locuaz.readthedocs.io/en/latest/tutorialsimple.html}
After the protocol run, all average scores had at least doubled and many promising candidates were recovered from the last epoch. These can then be further validated by longer MD simulations, rescored if necessary, and, finally, undergo experimental validation. 

\FloatBarrier 

\section{Competing interests}
No competing interest is declared.

\section{Acknowledgments}
PB acknowledges the TRIL fellowship for having provided support under the ICTP TRIL programme, Trieste, Italy.
The research leading to these results has received funding from AIRC under IG 2020 - project ID. 24589
We acknowledge PRACE for awarding us access to the supercomputing resources hosted by CINECA and
the VEGA resources hosted at the Institute of Information Science, IZUM.

\bibliographystyle{abbrvnat}
\bibliography{reference}

\end{document}